\documentclass[preprint,12pt]{elsarticle}

\usepackage{graphicx}
\usepackage{lineno}

\newcommand{\alt}{\mathrel{\mathop{\kern 0pt \rlap
  {\raise.2ex\hbox{$<$}}}
  \lower.9ex\hbox{\kern-.190em $\sim$}}}
\newcommand{\agt}{\mathrel{\mathop{\kern 0pt \rlap
  {\raise.2ex\hbox{$>$}}}
  \lower.9ex\hbox{\kern-.190em $\sim$}}}

\newcommand{\Saclay}{IRFU, Centre d'\'Etudes Nucl\'eaires de Saclay (CEA-Saclay), Gif-sur-Yvette, France}
\newcommand{\apc}{APC-Astroparticule et Cosmologie, CNRS-CEA-IN2P3-Observatoire de Paris, \\ \footnotesize \it Université Paris Diderot-Paris7, France}
\newcommand{\Zaragoza}{Laboratorio de F\'{\i}sica Nuclear y Astropartículas, Universidad de Zaragoza, Zaragoza, Spain }

\newcommand{\Zaragozax}{$^a$}
\newcommand{\Saclayx}{$^b$}
\newcommand{\apcx}{$^c$}

\begin{document}

%\linenumbers

\journal{Nuclear Instruments and Methods A}

\begin{frontmatter}

\title{Energy resolution of alpha particles in a microbulk Micromegas detector at high pressure Argon and Xenon mixtures}

\author{T. Dafni\Zaragozax, E. Ferrer-Ribas\Saclayx, I. Giomataris\Saclayx, Ph. Gorodetzky\apcx, \\F. Iguaz\Zaragozax, I. G. Irastorza\Zaragozax\footnote{corresponding author. E-mail: igor.irastorza@cern.ch. }, P.
Salin\apcx, A. Tom\'as\Zaragozax  }

\address{\Zaragozax\Zaragoza}

\address{\Saclayx\Saclay}
\address{\apcx\apc}

%\date{\today}

%\maketitle

\begin{abstract}

The latest Micromesh Gas Amplification Structures (Micromegas) are achieving outstanding energy
resolution for low energy photons, with values as low as 11\% FWHM for the 5.9 keV line of
$^{55}$Fe in argon/isobutane mixtures at atmospheric pressure. At higher energies (MeV scale),
these measurements are more complicated due to the difficulty in confining the events in the
chamber, although there is no fundamental reason why resolutions of 1\% FWHM or below could not be
reached. There is much motivation to demonstrate experimentally this fact in Xe mixtures due to the
possible application of Micromegas readouts to the Double Beta Decay search of $^{136}$Xe, or in
other experiments needing calorimetry and topology in the same detector. In this paper, we report
on systematic measurements of energy resolution with state-of-the-art Micromegas using a 5.5 MeV
alpha source in high pressure Ar/isobutane mixtures. Values as low as 1.8\% FWHM have been
obtained, with possible evidence that better resolutions are achievable. Similar measurements in
Xe, of which a preliminary result is also shown here, are under progress.

\end{abstract}


\end{frontmatter}

\section{Introduction}

The use of gas Time Projection Chambers (TPCs) as calorimeters,
like in recently explored applications of gamma ray astronomy,
dark matter or double beta decay, in contrast to their traditional
use as tracking detectors, has increased the interest of improving
their energy resolution and studying the factors that ultimately
limit it. The appearance of novel charge readout planes based on
micropattern techniques (like Micromegas) with substantially
better energy resolution capabilities, is a further motivation for
these studies.

In the specific case of Double Beta Decay (DBD) searches, the
Gothard TPC \cite{Luscher:1998sd} in the 90's represented a
pioneering use of a gaseous Xenon TPC to look for the neutrinoless
DBD of the $^{136}$Xe isotope. The modest energy resolution has
been, however, a drawback to further pursue this experimental
technique in the last decade. Indeed, for next generation
experiments aiming at sensitivities down to $\sim$100 eV or
$\sim$10 eV for the effective neutrino mass (for which target
masses of $\sim$100 kg or several tons are needed respectively),
energy resolutions better than 4.5\% and 2\% FWHM respectively at
the Q$_{\beta\beta}$ value are required \cite{Avignone:2005cs}.
These requirements on energy resolutions come from the need of
separating the tail of the standard DBD with 2 neutrinos, which
represent an irreducible background to the neutrinoless DBD. On
the other hand, it is well known that gas Xe TPCs can offer very
promising and unique features when compared with other
experimental techniques. Among other things, natural Xe already
contains 7\% of the required isotope and, as a gas, it is
relatively easy to enrich. In addition, the DBD event in gas has a
distinctive topological information that an appropriately
pixelised readout could exploit to identify the signal and reduce
backgrounds.

The advent of new readout techniques which may overcome the
limitations regarding energy resolution have renewed the interest
of gas Xe TPC for DBD searches. Recent proposals like the ones of
NEXT\cite{next} and EXO\cite{exo} collaborations are based on this
fact. The purpose of the present work is to evaluate which is the
best energy resolution of state-of-the-art Micromegas readout
planes.

The energy resolution in gaseous proportional counters (and by
extension in gaseous Time Projection Chambers with electron
avalanche readouts) depends on many factors. Some of them can be
considered non fundamental and can in principle be overcome
(although with difficulty in practice). Examples of these are
non-uniformity of readout planes, problems of equalization of
multiple channels, ballistic deficit, attachment to
electronegative impurities of the gas, time dependencies, etc...
The only truly fundamental effects limiting the energy resolution
are the fluctuations occurring in the number of primary
electron-ion pairs produced by the ionizing particle (and
determined by the Fano factor) as well as the fluctuations in the
number of secondary electrons produced in the avalanche initiated
by each primary electron.

\begin{figure*}[t]
\centering
\includegraphics[width=100mm]{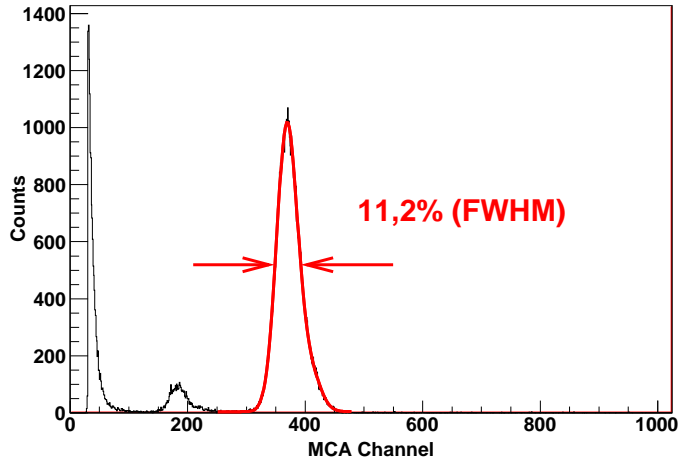}
 \caption{
Typical spectrum of $^{55}$Fe with a microbulk Micromegas. The red
line is the result of a fit to 2 gaussians.
 \label{55Fe}}
\end{figure*}

The Micromegas \cite{Giomataris:1995fq, Giomataris:2004aa} readouts make use of a metallic
micromesh suspended over a (usually pixellised) anode plane by means of insulator pillars, defining
an amplification gap of the order of 50 to 150 $\mu$m. Electrons drifting towards the readout, go
through the micromesh holes and trigger an avalanche inside the gap, inducing detectable signals
both in the anode pixels and in the mesh. It is known \cite{Giomataris:1998} that the way the
amplification develops in a Micromegas gap is such that its gain $G$ is less dependent on
geometrical factors (the gap size) or environmental ones (like the temperature or pressure of the
gas) than conventional multiwire planes or other types of micropattern detectors based on charge
amplification. This fact allows in general for higher time stability and spatial homogeneity in the
response of Micromegas, reducing the importance of some of the non-fundamental factors mentioned
above affecting the energy resolution. In addition, the amplification in the Micromegas gap has
less inherent statistical fluctuations than that of wires, due to the faster transition from the
drift field to the amplification field provided by the micromesh \cite{Alkhazov:1970fx}.
Experimentally, resolutions of 11\% FWHM for the 5.9 keV $^{55}$Fe peak, like the one shown in
figure \ref{55Fe} are achieved by the latest generation of Micromegas readouts.

Assuming a square root of energy dependency, this value would point to energy resolutions of less
then 1\% at the MeV scale, by far fulfilling the requirements, for instance, of double beta decay
applications \cite{Avignone:2005cs}. An experimental confirmation of this, however, is not an easy
task, due to the difficulty of confining the high energy events in a small chamber, and in fact
results on energy resolution in these conditions for gaseous TPCs are seldom found in the
literature.

In this paper we present the first results of our work aiming at
measuring the energy resolution at high energy and high pressure
of Micromegas readouts. In section \ref{csetup}, the experimental
setup used for the measurements is described in some detail. In
section \ref{argon_meas}, we present the methodology followed and
the main results achieved with argon/isobutane mixtures. In
sections \ref{q_meas} and \ref{xe_meas} some additional results on
the quenching factor in Argon/Methane mixtures and preliminary
results on pure Xenon are presented respectively. We finish with
our conclusions and future plans in section \ref{conclusions}.

\section{Experimental setup}\label{csetup}

\begin{figure*}[t]
\centering
\includegraphics[width=80mm]{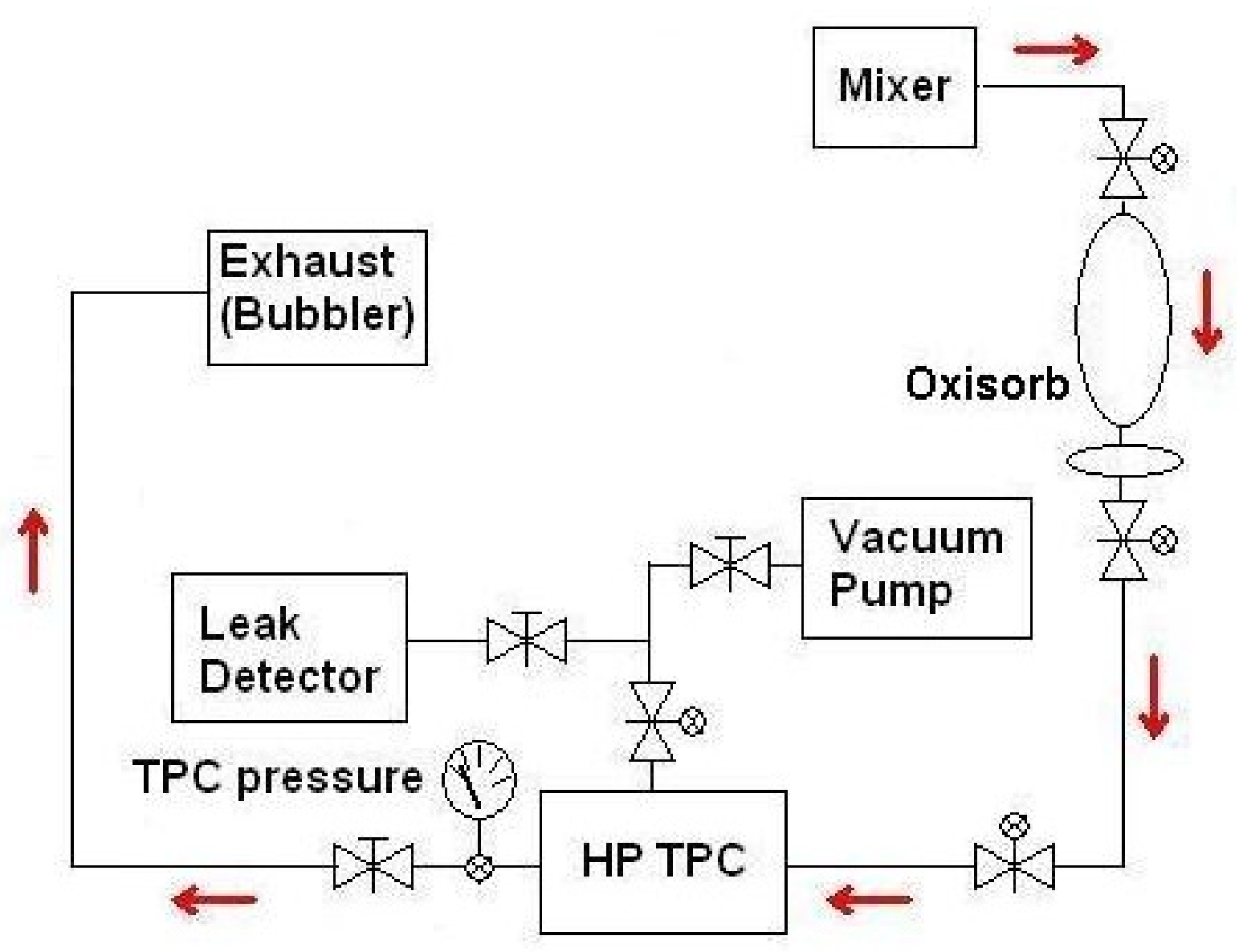}
 \caption{
Scheme of the setup.
 \label{setup}}
\end{figure*}

The experimental setup used in these measurements was adapted from
the former HELLAZ setup at CEA/Saclay\cite{Dolbeau:2005kt},
already designed to perform R\&D with custom-made gas mixtures of
high purity and at high pressures. The overall scheme of the
system is shown in figure \ref{setup}, and includes a high
pressure vessel of 1 liter of volume, a turbomolecular pump, a gas
distribution system with a gas mixer, an oxysorb filter, a
pressure controller, and an exhaust line with bubbler.

\begin{figure*}[t]
\centering
\includegraphics[width=75mm]{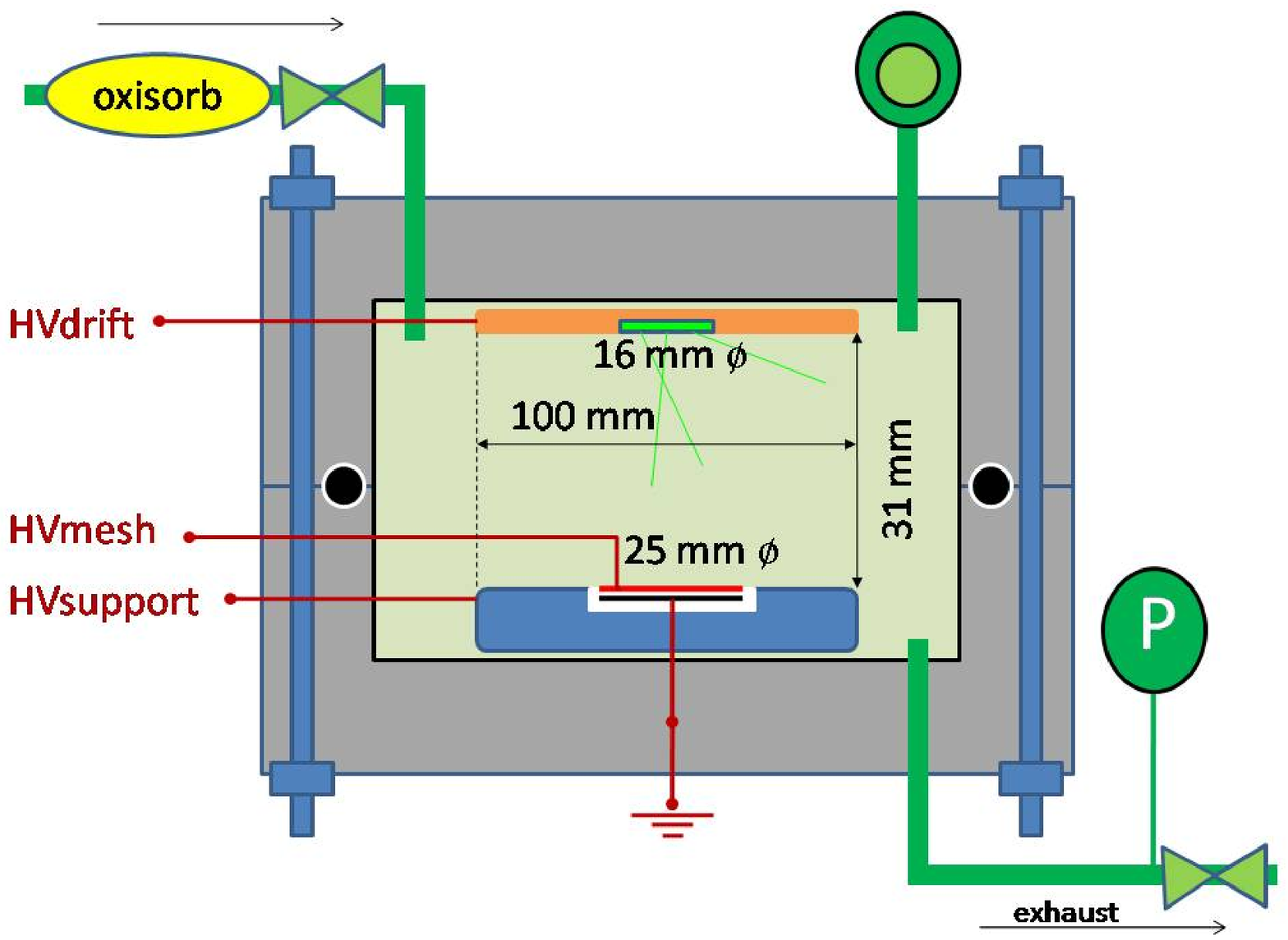}
\includegraphics[width=75mm]{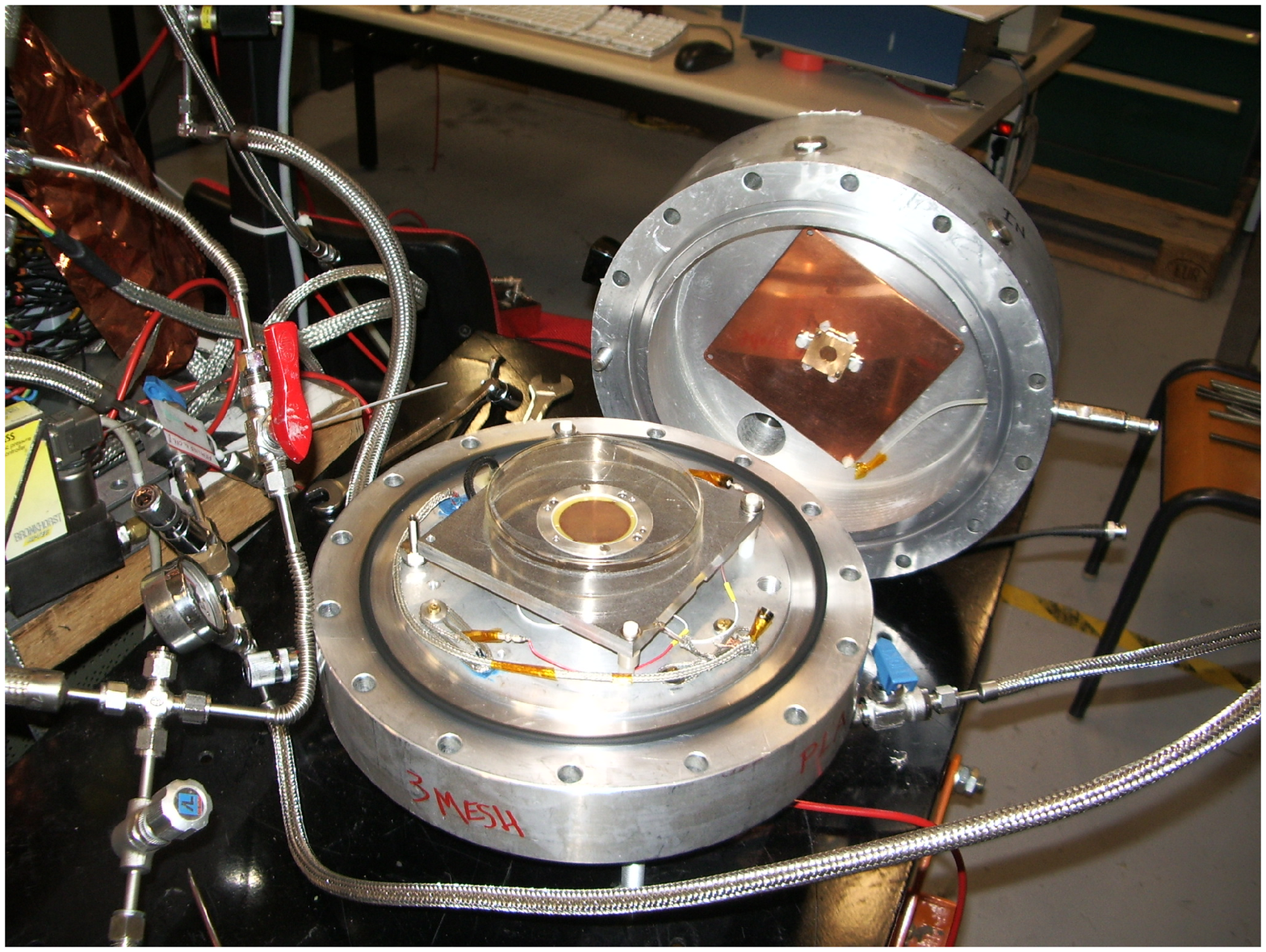}

 \caption{
On the left, schematic drawing of the setup inside the high pressure vessel, showing the position
of the Micromegas readout, the drift cathode and the source. On the right, photograph of the same
setup, showing the 2 vessel end-caps when open.
 \label{vessel}}
\end{figure*}

The main element of the setup is the high pressure vessel. It is
composed by two 5 cm thick aluminium caps sketched in fig
\ref{vessel}.a and photographed in \ref{vessel}.b, which close
with a viton o-ring and leave a 1-liter inner space able to hold
gas up to 20 bar \footnote{although in the present version of the
system a maximum pressure of 5 bar was respected by limitation of
other elements}. The dimensions of the inner space are 16 cm
diameter and 5 cm of height. The vessel is provided with
appropriate high voltage (SHV) and signal (BNC) feed-throughs for
the internal readout setup, as well as a CF40 outlet for pumping,
equipped with an all-metal UHV high-pressure valve. The gas mixer
is composed by 3 independent gas lines with Bronkhorst mass flow
controllers able to work at high pressure (20 bar). The gas
mixture passes through a MESSER oxysorb filter before entering the
chamber.

\begin{figure*}[t]
\centering
\includegraphics[width=100mm]{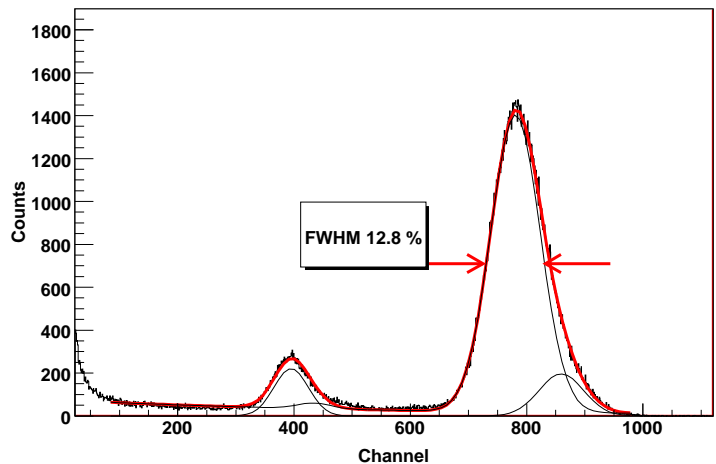}
 \caption{
$^{55}$Fe spectrum obtained with Ar + 5\% isobutane at atmospheric pressure with the same
Micromegas readout used in the measurements presented in this paper.
 \label{fe55_this}}
\end{figure*}

The inside of the high pressure vessel was equipped with a
state-of-the-art 2.3 cm diameter circular Micromegas readout of 50
microns gap, with a single non-segmented anode covering all this
area. The readout was made by the \emph{microbulk} manufacturing
technique developed at CERN\cite{microbulk}, which yields the
Micromegas amplification structures out of a double-clad kapton
sheet, by chemically removing part of the kapton. This technique
is known to yield the highest precision in the gap homogeneity
and, because of that, the best energy resolutions among
Micropattern detectors. The particular readout used in this
measurements was tested before and after the measurements campaign
with a $^{55}$Fe source at atmospheric pressure, presenting the
spectrum shown in fig. \ref{fe55_this}, with a resolution of
12.8\% FWHM at the 5.9 keV line of $^{55}$Fe. We must stress that
no appreciable deterioration was seen after the year-long duration
of the measurements campaign.

The readout was installed inside of the high pressure vessel by
means of a metallic support of $10 \times 10$ cm$^2$. Apart from
mechanically supporting the Micromegas readout, it aims at
extending the equipotential surface defined by the Micromegas mesh
and preserve a good shape of the drift electric field (i.e. drift
lines perpendicular to the Micromegas plane) all along the
conversion volume which projects onto the Micromegas surface. The
supporting piece is therefore placed at the same voltage as the
Micromegas mesh. The drift cathode is placed above the Micromegas
at a distance of 3.1 cm, and was prepared to hold the Americium
source in its center, as sketched in fig \ref{vessel}.

The electrical connections were made in such a way that we could
power independently the drift cathode, the Micromegas mesh and the
supporting piece. Although, as explained before, the voltage of
the supporting piece is usually set the same as the one of the
mesh in order to preserve the drift field. The signal was read out
from the Micromegas mesh using a CANBERRA 2004 preamplifier, whose
output was fed into an ORTEC VT120 amplifier/shaper and
subsequently into a multichannel analyzer AMPTEK MCA-8000A for
spectra building. Alternatively, the output of the preamplifier
was digitized directly by a LeCroy WR6050 oscilloscope and saved
into disk for further offline inspection.

\section{Measurements and results} \label{argon_meas}

The system was cleaned before every measurement either by pumping
it with the turbo-molecular pump down to pressures below 10$^{-6}$
mbar, or by purging it with argon several times. The purity of the
gas used was always laboratory-grade (4.6 or superior) provided by
MESSER. The final gas purity, determined in addition by the
leak-tightness of the system, the outgassing of materials and the
effect of the oxysorb, was proven to be adequate for the
short-term runs with Argon mixtures for a 3.1 cm drift distance
presented in this paper. For the forthcoming Xenon measurements
the present system may be limiting as evidence of some degree of
attachment was found. This is discussed in the section
\ref{xe_meas}.

The desired gas mixture was introduced into the high pressure
vessel using the gas mixer with the appropriate relative flows.
The gas mixtures for which data are presented in this section
include argon plus a variable amount of isobutane (between 2\% and
5\%). Methane was also used in some measurements for practical
reasons (non-availability of isobutane) although it is known not
to be an optimal mixture for Micromegas operation. The desired gas
pressure was set via the back-pressure controller. Two operation
modes have been used: 1) open-loop, i.e., the gas flow is directed
towards the exhaust line and lost, keeping the input flow at
typical values of a few l/h, and 2) sealed-mode, i.e. the vessel
was filled up to the desired pressure with the output outlet
closed, and then operation started with the gas sealed inside the
vessel. No substantial difference was observed between both modes
of operation in the Argon data.

All data here presented has been taken using an Americium-241
source. The source consisted on a metallic circular substrate of
25 mm diameter with the Americium deposited on its center, in an
approximate circular region of about 8 mm diameter. The source is
not sealed, that is, no material is present on top of the
americium that could stop the alphas. It was installed inside the
vessel, attached to the center of the drift cathode, and in
electrical contact with it, facing the center of the Micromegas
readout (see Fig. \ref{vessel}). In order to contain the 5.5 MeV
alpha particles from the main Am-241 line in our 3.1 cm of drift,
the gas needed to be at a pressure of at least $\sim 2$ bar. This
is the reason why no data are presented for atmospheric pressure.
The activity of the source was such that, when uncollimated, it
induced a typical rate of about 200 Hz on the Micromegas.

\begin{figure*}[t]
\centering
\includegraphics[width=100mm]{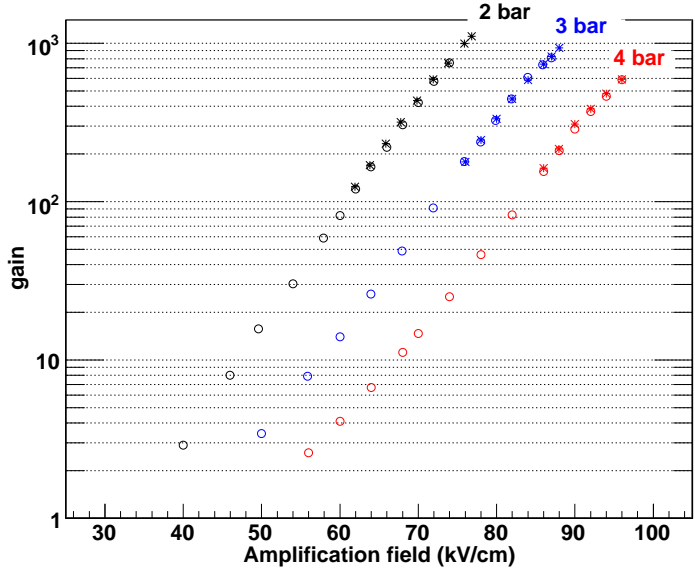}
 \caption{
Gain curves with both alpha (circles) and gamma (crosses) peaks,
for Ar+5\% isobutane, at 2, 3 and 4 bar.
 \label{qfactor}}
\end{figure*}

\begin{figure*}[p]
\centering
\includegraphics[width=100mm]{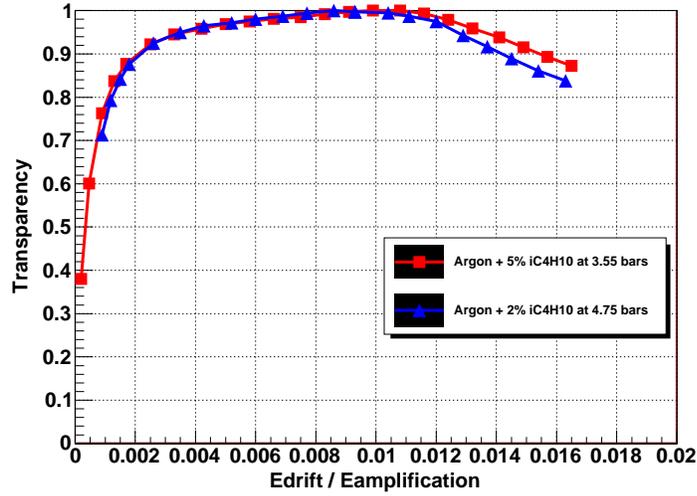}
 \caption{
Dependence of the alpha peak position on the drift/amplification
field ratio (\emph{electron transmission curve}). Values are
normalized to the maximum of each series.
 \label{transparency}}
\end{figure*}

The typical sequence of data-taking for a given gas mixture and
pressure included systematic variation of the mesh and drift
voltages, registering the position of the 5.5 MeV alpha peak, as
well as its width. The mesh voltage is typically varied in the
range from 200 V to 500 V, corresponding to amplification fields
from 40 to 100 kV/cm. The specific values for each pressure can be
seen in fig. \ref{qfactor}. The drift voltage is typically varied
from 300 V up to 4000 V (limit of the power supply used)
corresponding to drift fields from 100 V/cm to 1333 V/cm. The
evolution of the peak position versus the mesh voltage gave the
typical gain curves like the ones shown in fig. \ref{qfactor}. The
evolution of the peak position versus the drift field (for a fixed
mesh voltage) showed the characteristic flat dependence shown in
Fig. \ref{transparency} over a wide range of drift/amplification
field ratios (0.002 -- 0.012), decreasing for larger drift fields
due to decrease of the electron transmission of the mesh. The
curve is compatible with usual curves taken with the $^{55}$Fe
source at atmospheric pressure and shows that no significant
attachment is present in the system. The relevant data were taken
in conditions corresponding to the full electron transmission
through the mesh (on the plateau of the curve \ref{transparency}).

%
%\begin{figure*}[t]
%\centering
%\includegraphics[width=120mm]{ArgonGain.eps}
% \caption{
%Gain curves obtained at different pressures (from 2 to 5 bars) in
%argon/isobutane mixtures (2\% and 5\%). Voltages were not raised
%until the highest possible values without sparking, so the maxima
%of the plots do not correspond to the highest achievable gains as
%is often customary in this kind of plots.
% \label{gain}}
%\end{figure*}

\begin{figure*}[p]
\centering
\includegraphics[width=75mm]{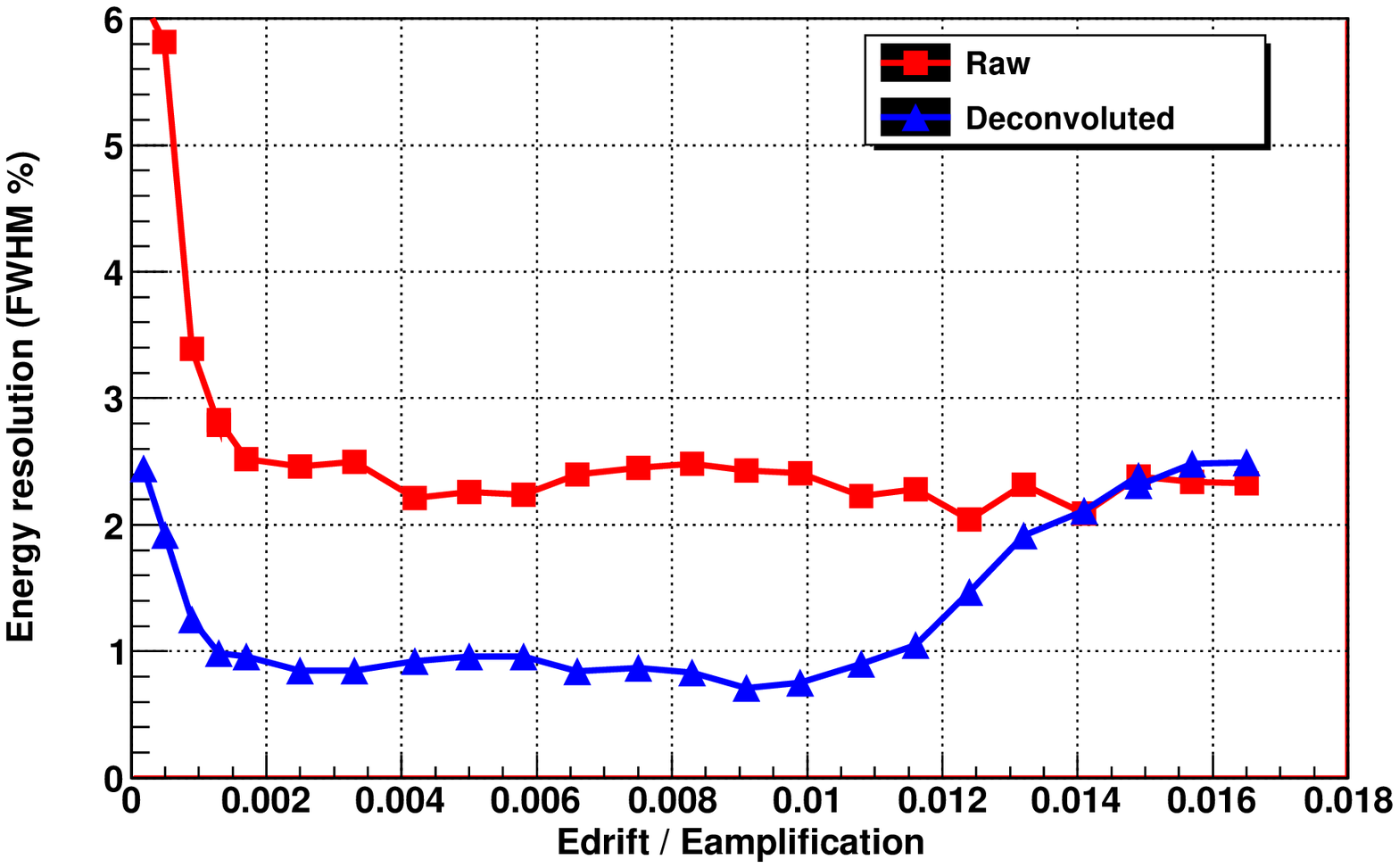}
\includegraphics[width=75mm]{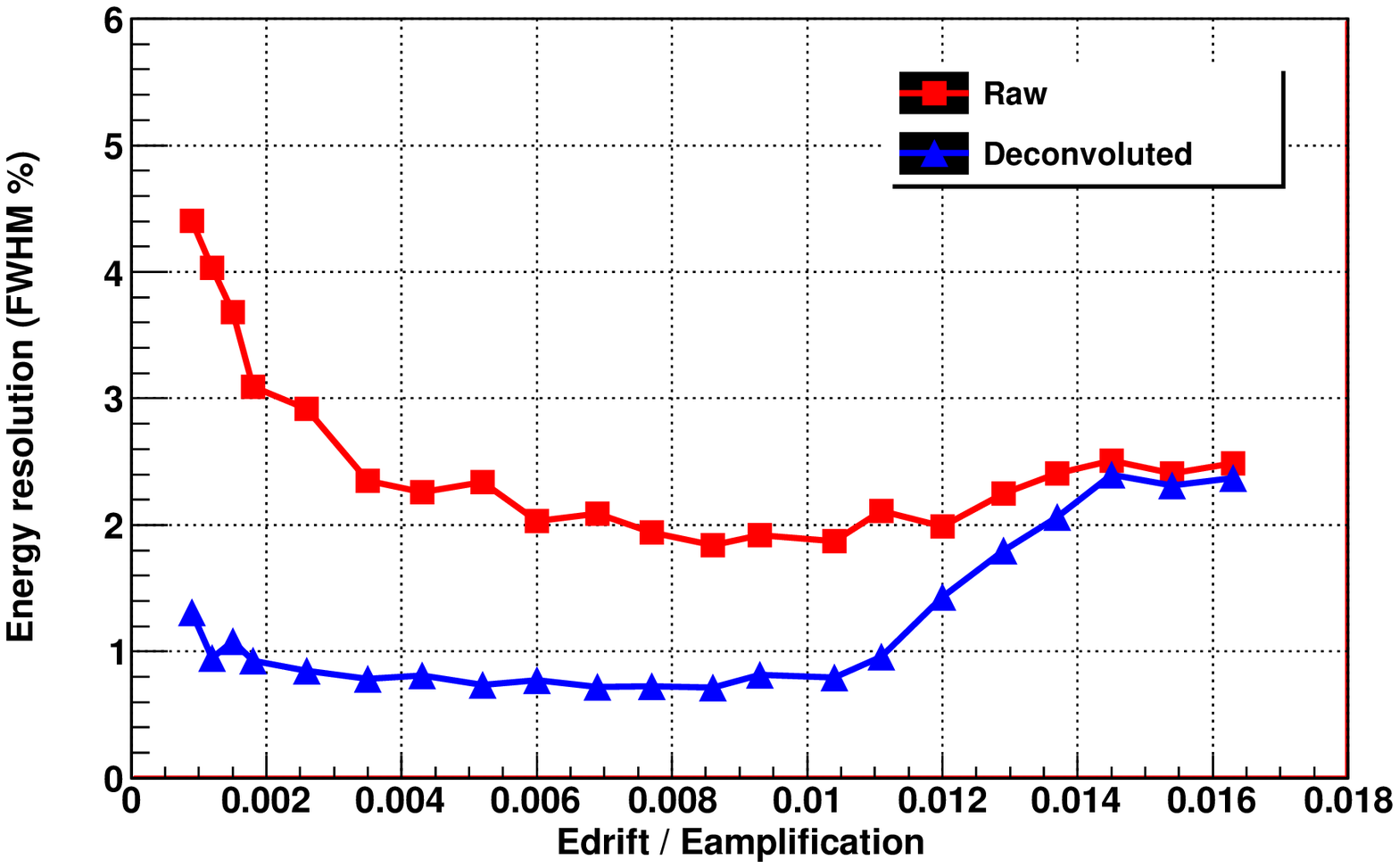}
 \caption{
Dependence of the energy resolution measurements on the drift/amplification field ratio at 4 bar
for both the 2\%-isobutane (left) and the 5\%-isobutane cases. Squares correspond to the FWHM of
the peak obtained by simple fit to gaussian, while triangles correspond to the deconvoluted
resolution obtained by fitting to gaussian+landau as described in the text.
 \label{res_vs_drift}}
\end{figure*}

The FWHM of the alpha peak shows a characteristic dependency on
the drift field, shown in Fig.~\ref{res_vs_drift}, clearly
correlated with the electron transmission plot of
Fig.~\ref{transparency}. Best energy resolutions are obtained for
larger drift fields inside the electron transmission plateau,
although resolutions around the level of 2\% and 2.3\%
(respectively for 2\% and 5\% isobutane mixtures) are obtained for
a relatively wide range of values. The best obtained values were
1.8\% and 2\% respectively.

\begin{figure*}[t]
\centering
\includegraphics[width=100mm]{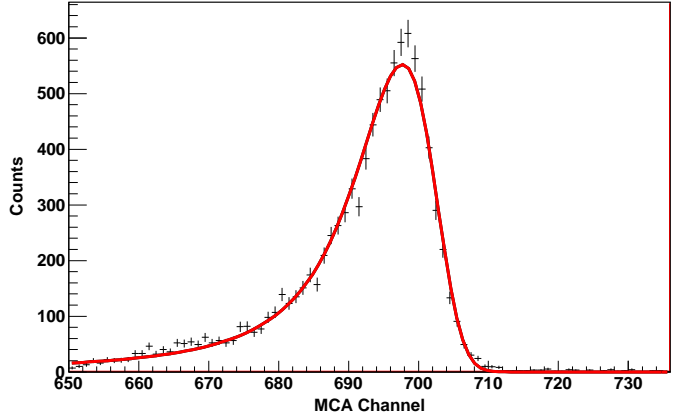}
 \caption{
Example of alpha peak measured, with the fit to a Landau function
convoluted with a gaussian. The best fit values for the FWHM of
the gaussian component is 0.7\%.
 \label{peak}}
\end{figure*}

In the best cases, the peak showed a clearly asymmetric shape as
it is shown in Fig.~\ref{peak}. As this shape is not expected from
just fundamental fluctuations in the primary or secondary charge,
it is indicative of (small) external effects related, for example,
to incomplete charge collection for some events (for example for
alphas coming from the bulk of the Americium and leaving some
energy in it before exiting, or cases where the alpha tracks are
in small angle with the cathode plane and part of its ionization
is lost back to the cathode). Let us add that this asymmetry
cannot be attributed to the effect of attachment, as the voltage
configurations for which it is observed lie always well within the
electron transmission plateau. In any case the amount of charge
loss would very plausibly follow a Landau distribution. In fact,
the peak shape is very well parameterized by a negative Landau
function convoluted by a gaussian. If we attribute the mentioned
external effects to the Landau function we can see the remaining
gaussian as the intrinsic energy resolution of the Micromegas.
This deconvolution analysis, illustrated in the fit of
Fig.~\ref{peak}, points to an energy resolution of 0.7\% FWHM
(0.9\% for the 5\% isobutane case). This value could be suggestive
of the achievable resolution in a setup where the mentioned
external effects were not present. Measurements with a gaseous
source (radon) are envisaged to test the suggested origins of such
effects.

It is interesting to note the dependence of this "deconvoluted"
resolution with drift/amplification field ratios, shown in Fig.
\ref{res_vs_drift}. The values obtained are rather constant along
the electron transmission plateau and get worse beyond it (the
peak becomes symmetric). It is reasonable to assume that whatever
the origins of the peak assymetry, they just become less important
compared with other (symmetric) effects of gain fluctuation at
higher field ratios (electron loss in the mesh, larger statistical
fluctuations in avalanche, etc…).

In summary, energy resolution at the 2\% FWHM or below have been
demonstrated in 4 bar argon/isobutane mixtures for the 5.5 MeV
alpha line. Possible evidence for resolutions below 1\% is pointed
out based on the asymmetry of the peak, although further work is
needed to identify the effects causing the asymmetry, and
eventually remove them.

\begin{table}[t]
\begin{center}
\begin{tabular}{ccc}
\hline Pressure (bar) & Gas Mixture & Energy resolution (FWHM)\\
\hline
%%2 & Ar + 5\% isobutane & 3.0 \% \\
4 & Ar + 2\% isobutane & 1.8 \% \\
4 & Ar + 5\% isobutane & 2.0 \% \\
 \hline
\end{tabular}
\caption{Best energy resolutions achieved for each argon-isobutane
mixture at 4 bar of pressure.\label{table}}
\end{center}
\end{table}

\section{Alpha ionization yield results}  \label{q_meas}

Although the average energy required to ionize (the $W$ value) is
similar for alphas or electrons \cite{Knoll}, the effective
ionization of an alpha particle of a given energy may be less than
that of an electron of the same energy in some conditions. This is
due to the effect of recombination of electron-ion pairs, which is
stronger in the denser alpha tracks. This effect is dependent on
the density of the media (being more relevant, for example, in
noble liquid detectors) and on the applied electric drift field
(strong electric fields prevents the recombination). The presence
of quenchers could also have some effect.

\begin{figure*}[p]
\centering
\includegraphics[width=100mm]{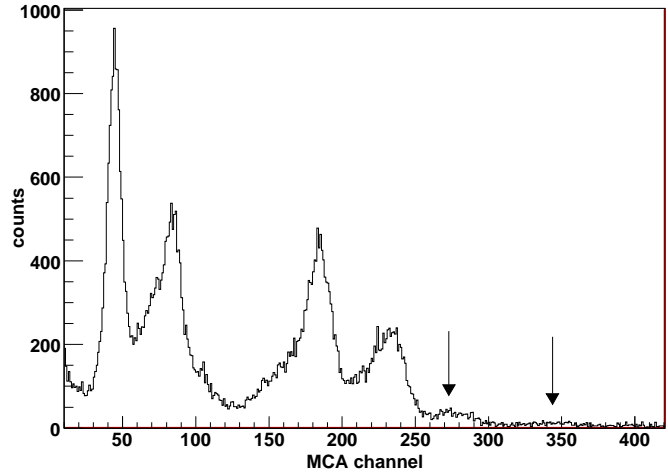}
 \caption{
Spectrum of low the $^{241}$Am low energy gamma lines, obtained
with the alphas blocked: 3.0 keV and 6.1 keV fluorescences; 13.9
keV, 17.7 keV and 21.0 keV X-rays; 26.4 keV gamma lines from
$^{241}$Am
 \label{xrays_clean}}
\end{figure*}

In the present section we gather the results achieved to determine whether in our specific
conditions alphas suffer from any appreciable recombination, in which case a correction, or
\emph{quenching factor}, would be needed to apply in order to know which electron-equivalent energy
corresponds our 5.5 MeV alpha peak. The method followed to measure the alpha quenching factor was
to focus on the x-rays and gamma lines of the same Americium source. In order to see these lines,
of considerable lower energy than the alpha one, we started by blocking the source with a thin
aluminium foil that would stop the alpha emission. The Micromegas mesh was set at higher voltages
with respect to the alpha runs. The result is illustrated in Fig. \ref{xrays_clean}, where a series
of peaks are identified (see caption). We proceed to calculate the gain curve using these peaks,
checking that all gave identical curves, as expected from gain proportionality. Finally, we compare
the curve obtained with the gamma peaks to that of the alpha peak.

\begin{figure*}[p]
\centering
\includegraphics[width=100mm]{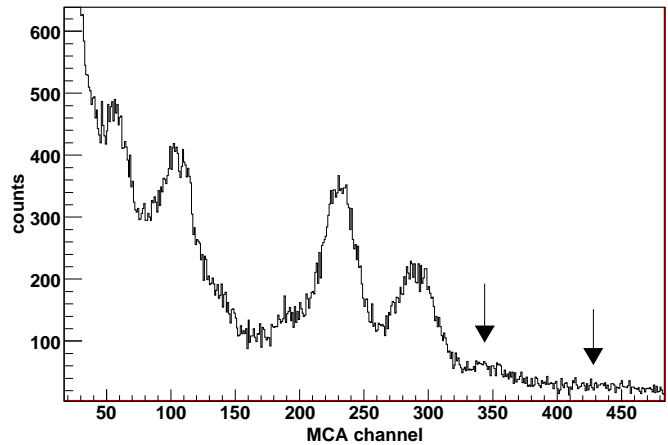}
 \caption{
Same spectrum of energy lines of $^{241}$Am over a background continuum associated to the alphas.
\label{xrays_background}}
\end{figure*}

The need of covering and uncovering the alpha source (and opening the system, re-pumping, and
re-filling) left doubts on the reproducibility of the gas conditions from measurement to
measurement. In order to obtain a reliable result the alpha peak should be observed in the same
runs as the gamma peaks. For that reason we collimated the alpha source in such a way that a small
alpha peak was observable without contaminating too much the low energy region and thus still
observe the x-ray and gamma peaks. Fig. \ref{xrays_background} shows the low energy spectrum
obtained in such conditions.

The gain curves for both the alpha and the gamma peaks are shown in Fig. \ref{qfactor}. As can be
seen both kind of lines are coincident and therefore the quenching factor is almost unity. The
precise values obtained by averaging over all the series are listed in table \ref{qtable} for 2, 3
and 4 bar. The error shown in Table \ref{qtable} represents only the 1 sigma statistical error from
the dispersion of the points of the averaged series and is therefore underestimated. From this we
conclude that recombination effects are very small in the experimental cases tested. Similar
measurements were performed in Ar + CH$_4$ mixtures (with different amounts of CH$_4$) with similar
results.

\begin{figure*}[t]
\centering
\includegraphics[width=100mm]{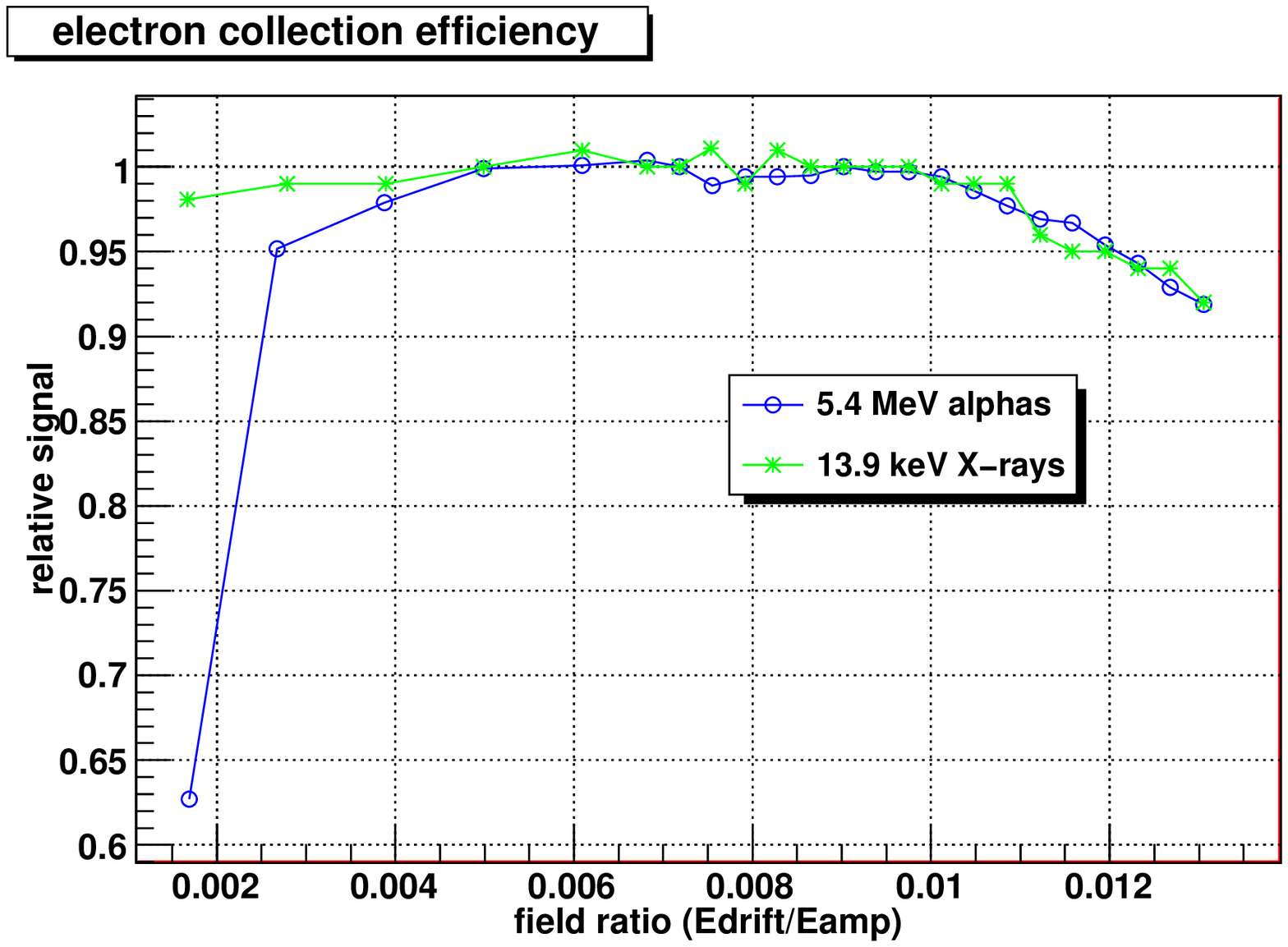}
 \caption{ Electron transmission curves for both X-rays (crosses) and alphas (circles) for Ar + 5\% isobutane at 4 bar of
pressure. Both curves
 are compatible but for the points at the lowest field ratios (lowest drift field values) for which
 alphas provide lower relative signals than X-rays. This could be attributed to the appearance of
 recombination effects at such low electric fields.
 \label{recomb}}
\end{figure*}

It is worth noticing that all these data were taken with values of drift sufficiently high so that
we were well within the plateau of Fig. \ref{transparency}. Additional data were taken at different
values of the drift field, as shown in Fig. \ref{recomb}, and only at very low values we measured
decreased values of the alpha series with respect to the gamma series, what can be attributed to
recombination effects.

\begin{table}[t]
\begin{center}
\begin{tabular}{ccc}
\hline Gas & Pressure (bar) & Q$_\alpha \pm $ 1$\sigma$ error\\
\hline
 Ar + 5\% isobutane & 2 & 0.987 $\pm$       0.006  \\
 Ar + 5\% isobutane & 3 &  0.984   $\pm$     0.007  \\
 Ar + 5\% isobutane & 4 &  0.966      $\pm$  0.010 \\
 \hline
\end{tabular}
\caption{Obtained values for the alpha quenching factor for Ar + 5\% isobutane at 2, 3 and 4 bar of
pressure. The error shown represents only the 1 sigma statistical error from the dispersion of the
points of the averaged series and is therefore underestimated. \label{qtable}}
\end{center}
\end{table}

\section{Preliminary results in pure Xenon}\label{xe_meas}

In this section we present our very preliminary results of
measuring the energy resolution in pure Xenon. As will be
discussed later the present status of the system is not suitable
to achieve the required levels of gas purity to work with Xenon,
and therefore the results here achieved are to be considered as
preliminary and certainly a conservative upper limit to the energy
resolution achievable.

\begin{figure*}[p]
\centering
\includegraphics[width=100mm]{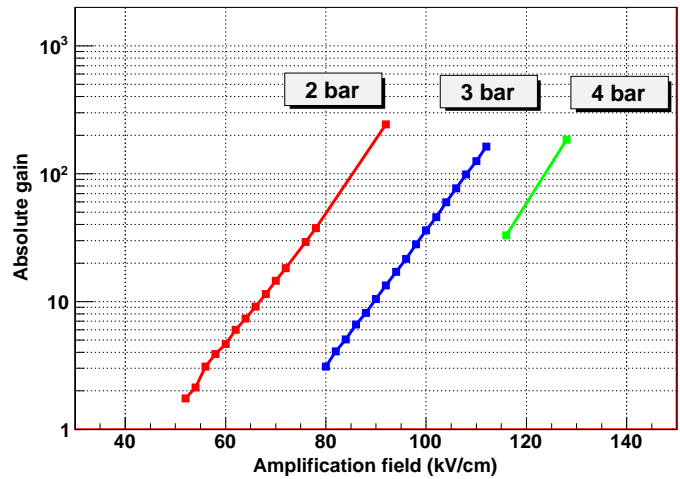}
 \caption{
Gain curves for pure Xenon for different pressures. Maxima correspond to the last point after which
instabilities and sparking begun.
 \label{gainXe}}
\end{figure*}

The methodology followed in this case is identical to the one
sketched in section 2, but only sealed-mode was used. The mesh
voltages used were somehow higher than the case of Argon, ranging
from 250 V to 650 V, corresponding to 50 to 130 kV/cm). The
specific values for each pressure can be seen in fig.
\ref{gainXe}. As in the case of Argon, gain and electron
transmission curves were measured (see Fig. \ref{gainXe}), for gas
pressures 2, 3 and 4 bar. A remarkable result was that the same
maximum gain of at least 200 \footnote{value obtained assuming a
quenching factor unity, so it is a lower limit} was obtained
independently of the pressure, indicating that this was a
limitation probably related to the amount of charge generated
locally in the Micromegas gap, set by the rate of alphas of our
source, and not the intrinsic Micromegas limit. Single runs with
the source more collimated showed indeed higher gains. We cannot
exclude that the same impurities that produce attachment (see
later) are allowing us to obtain larger gains, but it could also
be that the particular microbulk geometry is able to reach
relatively high gains in pure Xenon compared to other
amplification geometries. This question needs further study.

\begin{figure*}[p]
\centering
\includegraphics[width=100mm]{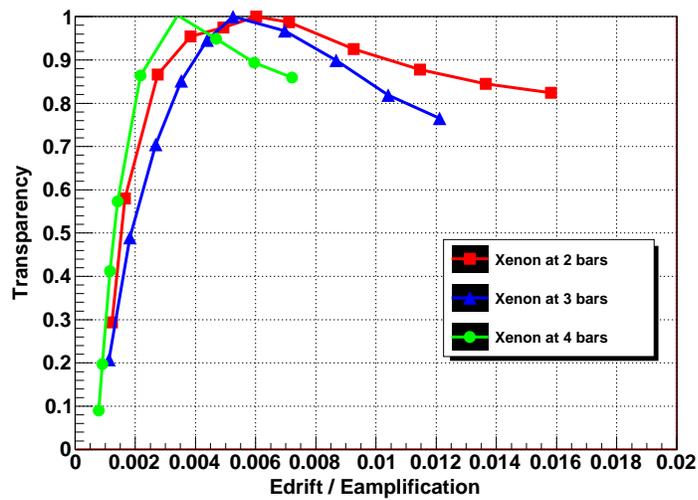}
 \caption{
Electron transmission  curves for pure Xenon at different
pressures. Values normalized to the maxima of each series.
 \label{transXe}}
\end{figure*}

\begin{figure*}[t]
\centering
\includegraphics[width=100mm]{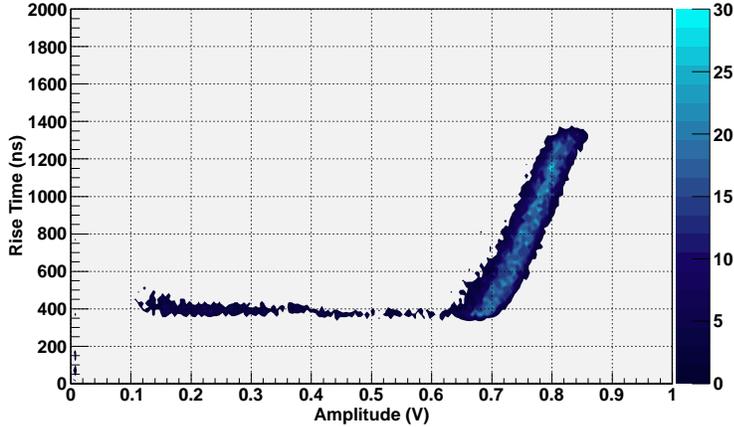}
 \caption{
Histogram of risetime and amplitude of alpha pulses taken at 4 bars of pure Xenon. It shows a clear
correlation between both quantities due to attachment, as explained in the text.
 \label{attachment}}
\end{figure*}

\begin{figure*}[t]
\centering
\includegraphics[width=75mm]{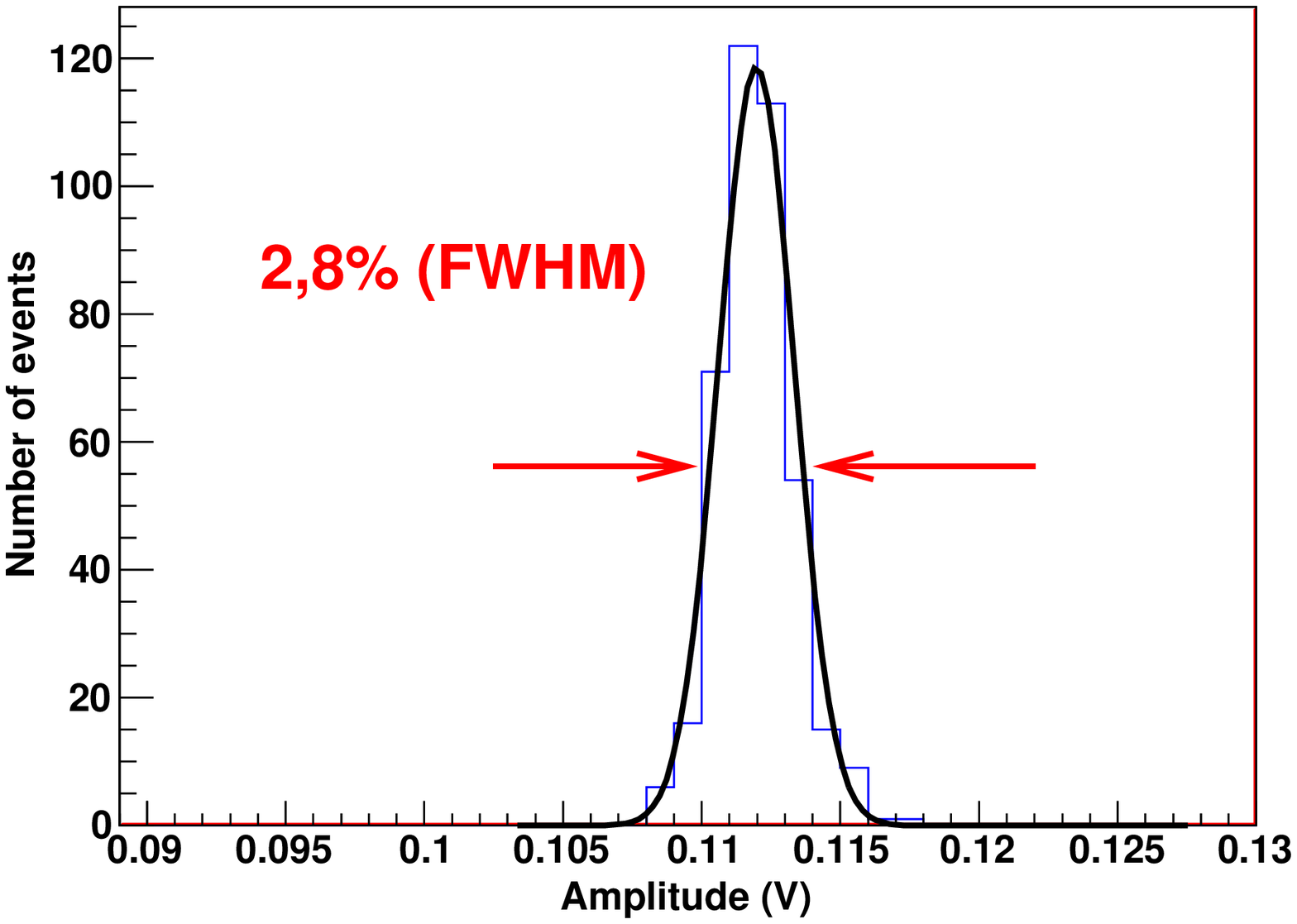}
\includegraphics[width=75mm]{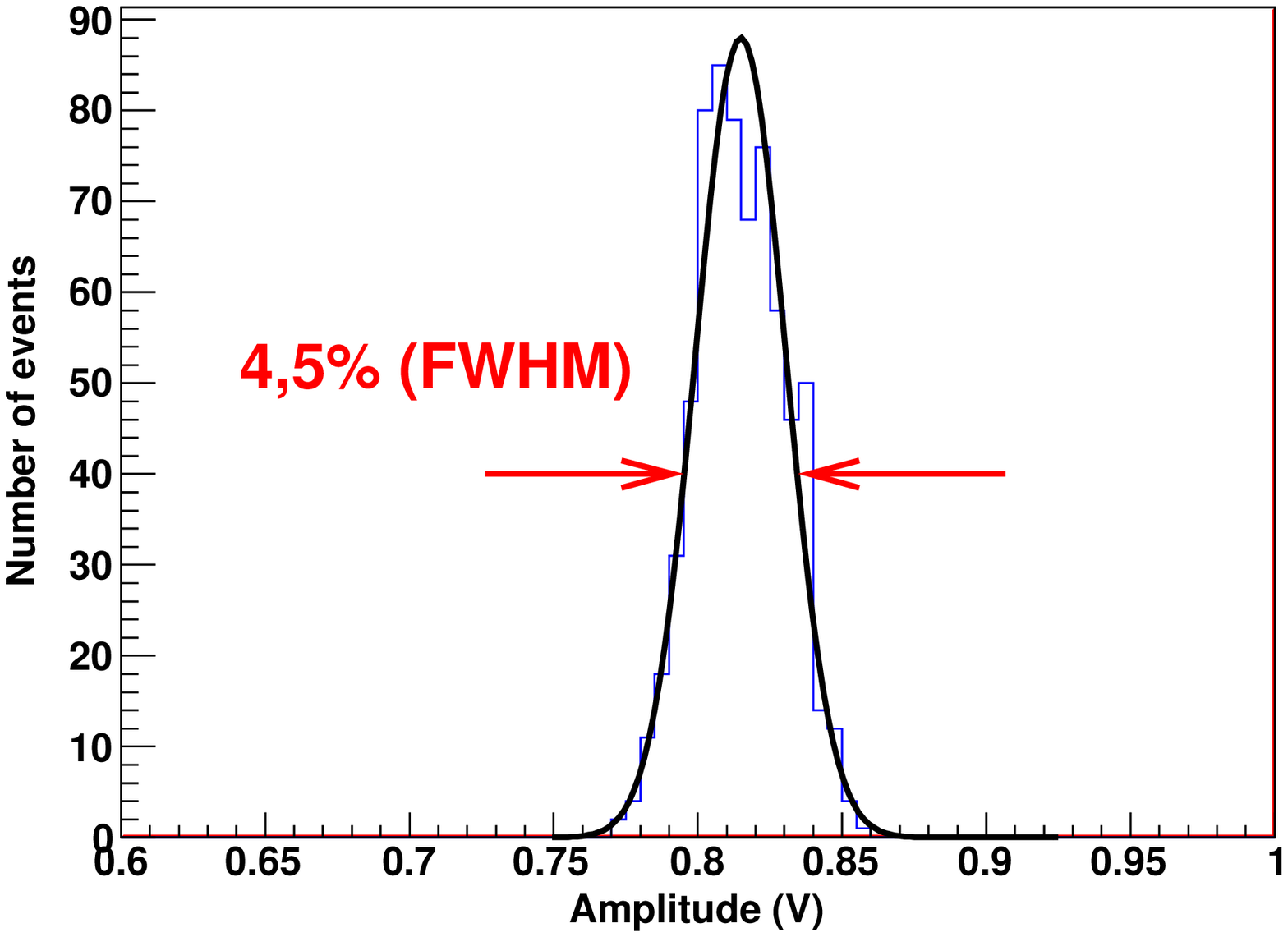}
 \caption{
Spectra of the alpha peak in pure Xenon at 2 bar (left) and 4 bar
(right), for events passing the cut on risetime described in the
text.
 \label{resXe}}
\end{figure*}

The curves obtained (Fig. \ref{transXe}) for the electron
transmission through the mesh showed evidence of attachment,
explaining why spectra taken with the MCA showed always modest
energy resolution. Data taken with the oscilloscope, allowed us to
perform a pulse shape analysis (PSA) to extract pulse
characteristics like risetime and amplitude. These data, unlike
the Argon ones, showed a clear correlation between the risetime of
the pulse and its amplitude, for events in the alpha peak, as
illustrated in Fig. \ref{attachment}. Pulses of long risetime
correspond to alpha tracks with orientation close to perpendicular
to the Micromegas plane, while pulses of shorter risetime
corresponds to tracks closer to parallel to the Micromegas plane.
The fact that the latter have lower amplitudes is an evidence of
attachment. Nevertheless the PSA allows us to have a conservative
estimation of the energy resolution without attachment by
performing a cut on risetime and keeping only events with similar
risetime, and therefore minimal spread due to different
attachments. The choice of risetime values for the plots here
presented (fig. \ref{resXe}) is 1800-2000 ns for the 2 bar case
and 1200-1300 ns for the 4 bar case. The values obtained with this
analysis were 2.8\% FWHM for pure Xenon at 2 bar, and 4.5\% FWHM
for pure Xenon at 4 bar, as shown in Fig. \ref{resXe}. We must
stress that these numbers are conservative, because the subset of
events kept by the cuts suffer still from attachment.
Nevertheless, the values obtained are positive and already
approach the requirements of double beta decay experiments. At
present we work to improve the system from the point of view of
leak-tightness and outgassing in order to repeat these measurement
in better conditions.

This is the first time that energy resolution measurements of
alpha particles at high pressure with microbulk type of Micromegas
are published. Similar measurements were done in
\cite{Ounalli:2008ue} for Xe-CF${_4}$ gas mixtures, although with
considerable worse values for the energy resolution. The reason
seems to be the type of readout used (woven-mesh) of significantly
less precision in the microscopic dimensions of the amplification
structure. This points to very promising prospects for the
application of these new type of readouts in calorimetric
experiments, and more specifically to double beta decay.

\section{Conclusions and prospects}\label{conclusions}

We have presented first results from measurements of energy resolution at high energy and high
pressures of Micromegas readouts. The main result regards measurements in argon/isobutane mixtures
at 4 bar. Energy resolution down to 1.8\% and 2\% FWHM at the $^{241}$Am 5.5 MeV alpha line have
been obtained for a wide range of drift/amplification field ratios, for percentages of isobutane of
both 2\% and 5\% respectively. Possible evidence for better achievable resolutions (down to 0.7\%
and 0.9\% FWHM respectively) are pointed out based on the asymmetry of the measured alpha peak.
Complementary to these measurements, it has been determined that the quenching factor of alpha
particles in argon + 5\% isobutane is almost unity, by comparing x-ray and gamma peaks of the
$^{241}$Am source.

Finally, preliminary measurements in pure Xenon have been
presented, although the status of the system have been proved not
to be good enough for electron drifting in Xenon, and effects of
some attachment in the gas were found. Nevertheless, by performing
pulse shape analysis on the digitized waveforms of the
preamplifier outputs and performing cuts on the risetime, a
conservative estimation of the energy resolution of 2.8\% FWHM for
2 bar and 4.5\% FWHM for 4 bar were found.

This is the first time that energy resolution measurements of
alpha particles at high pressure with microbulk type of Micromegas
are published. Even preliminary, the obtained energy resolution in
pure Xenon approaches the requirements of double beta decay,
opening very promising prospects for their use in these kind of
experiments.

At present we are working on the upgrade of the system to improve
its leak-tightness and outgassing in order to resume the
measurements in pure Xenon in appropriate conditions of gas
purity. We are also implementing a suitable recirculation system
with adequate filters to keep the gas purity for longer periods.

We also plan to use an enlarged high pressure vessel big enough to contain high energy electron
events, and perform similar measurements with photon/electron sources and not only with alphas.
Furthermore, we intend to explore the effect of quenchers in Xenon in the energy resolution as well
as going up to pressures higher than the ones in the present paper (from 5 to 10 bars).

Finally, we are also working on different enhancements of the
readouts that could lead to improved working conditions in high
pressure. For example, meshes with larger transparencies that the
one used here would allow us to work at higher drift fields and
partially overcome the effect of attachment that could jeopardize
the detector performance at large drift distances. In addition,
Micromegas with smaller amplification gaps should provide enhanced
stability of operation and therefore better energy resolutions
when operated at high pressure by virtue of the argument mentioned
in the introduction and described in \cite{Giomataris:1998}.
Amplification gaps below 50 microns pose new technical questions
for the manufacturing techniques that are currently being
addressed successfully. In particular, microbulks with gaps of 25
microns have recently being manufactured, and even smaller gaps
are planned. We plan to test them in the context of the present
line of work.

\section{Acknowledgements}

We wish to thank the staff of CEA/Saclay for the technical support received, in particular to S.
Aune and J.P. Mols. We also want to thank colleagues from the CAST, HELLAZ and NEXT collaborations
for many fruitful discussions. We acknowledge partial support from project FPA2007-62833 of the
Spanish Science and Innovation Ministry (MCINN). F. I. and A. T. acknowledge support from grants
AP2005-0360 of the MCINN and B024/2006 from the Aragon Regional Government.

%%%%%%%%%%%%%%%%%%%%%%%%%%%%%%%%%%%%%%%%%%%%%%%%%%%%%%%%%%%%%%%%%%%%%%%%%
%%%%%%%%%%%%%%%%%%%%%%%%%%%%%%%%%%%%%%%%%%%%%%%%%%%%%%%%%%%%%%%%%%%%%%%%%


\begin{thebibliography}{100}

%\cite{Luscher:1998sd}
\bibitem{Luscher:1998sd}
  R.~Luscher {\it et al.},
  %``Search for beta beta decay in Xe-136: New results from the Gotthard
  %experiment,''
  Phys.\ Lett.\  B {\bf 434} (1998) 407.
  %%CITATION = PHLTA,B434,407;%%


%\cite{Avignone:2005cs}
\bibitem{Avignone:2005cs}
  F.~T.~Avignone, G.~S.~King and Yu.~G.~Zdesenko,
  %``Next generation double-beta decay experiments: Metrics for their
  %evaluation,''
  New J.\ Phys.\  {\bf 7} (2005) 6.
  %%CITATION = NJOPF,7,6;%%

%\cite{Giomataris:1995fq}
\bibitem{Giomataris:1995fq}
  Y.~Giomataris, P.~Rebourgeard, J.~P.~Robert and G.~Charpak,
  %``MICROMEGAS: A high-granularity position-sensitive gaseous detector for
  %high particle-flux environments,''
  Nucl.\ Instrum.\ Meth.\  A {\bf 376} (1996) 29.
  %%CITATION = NUIMA,A376,29;%%

%\cite{Giomataris:2004aa}
\bibitem{Giomataris:2004aa}
  I.~Giomataris {\it et al.},
  %``Micromegas in a bulk,''
  Nucl.\ Instrum.\ Meth.\  A {\bf 560} (2006) 405
  [arXiv:physics/0501003].
  %%CITATION = NUIMA,A560,405;%%

\bibitem{Giomataris:1998}
  I.~Giomataris {\it et al.},
  Nucl.\ Instrum.\ Meth.\  A {\bf 419} (1998) 239.


%\cite{Alkhazov:1970fx}
\bibitem{Alkhazov:1970fx}
  G.~D.~Alkhazov,
  %``Statistics of electron avalanches and ultimate resolution of proportional
  %counters,''
  Nucl.\ Instrum.\ Meth.\  {\bf 89} (1970) 155.
  %%CITATION = NUIMA,89,155;%%


%\cite{Dolbeau:2005kt}
\bibitem{Dolbeau:2005kt}
  J.~Dolbeau, I.~Giomataris, P.~Gorodetzky, T.~Patzak, P.~Salin and A.~Sarat,
  %``The solar neutrino HELLAZ project,''
  Nucl.\ Phys.\ Proc.\ Suppl.\  {\bf 138} (2005) 94.
  %%CITATION = NUPHZ,138,94;%%

%\cite{Ounalli:2008ue}
\bibitem{Ounalli:2008ue}
  L.~Ounalli, J.~L.~Vuilleumier, D.~Schenker and J.~M.~Vuilleumier,
  %``New gas mixtures suitable for rare event detection using a Micromegas-TPC
  %detector,''
  JINST {\bf 4} (2009) P01001
  [arXiv:0810.0445 [hep-ex]].
  %%CITATION = JINST,4,P01001;%%

\bibitem{next}
J.~Díaz on behalf of NEXT collaboration, talk given at the Fourth Symposium on Large TPCs for Low
Energy Rare Event Detection, December 18th-19th, 2008, Paris. http://www-tpc-paris.cea.fr/

\bibitem{exo}
D. Sinclair on behalf of EXO collaboration, talk given at the Fourth Symposium on Large TPCs for
Low Energy Rare Event Detection, December 18th-19th, 2008, Paris. http://www-tpc-paris.cea.fr/


\bibitem{microbulk}
T. Papaevangelou, talk given at the Fourth Symposium on Large TPCs for Low Energy Rare Event
Detection, December 18th-19th, 2008, Paris. http://www-tpc-paris.cea.fr/

\bibitem{Knoll} Glenn F. Knoll. \emph{Radiation Detection
and Measurement.} 3rd edition. John Wiley and Sons. 2000.


\end{thebibliography}
\end{document}